\documentclass[12pt,a4paper]{article}
\usepackage{cite}
\usepackage{amsmath,amssymb,graphicx}
\usepackage{float}

\numberwithin{equation}{section}
\allowdisplaybreaks[3]

\newcommand{\al}{\alpha}
\newcommand{\ep}{\epsilon}

\newcommand{\lb}{\lbrack}
\newcommand{\rb}{\rbrack}

\newcommand{\msc}[1]{\mbox{\scriptsize #1}}
\newcommand{\dsp}{\displaystyle}
\newcommand{\scs}[1]{{\scriptstyle #1}}

\newcommand{\bm}[1]{{\boldmath #1}}

\newcommand{\bc}{\Bbb C}
\newcommand{\br}{\Bbb R}
\newcommand{\bz}{\Bbb Z}

\newcommand{\bh}{\Bbb H}

\newcommand{\bsz}{\Bbb Z}

\newcommand{\cJ}{{\cal J}}

\newcommand{\cN}{{\cal N}}
\newcommand{\cM}{{\cal M}}
\newcommand{\cF}{{\cal F}}

\newcommand{\cQ}{{\cal Q}}

\newcommand{\tL}{\tilde{L}}
\newcommand{\tJ}{\tilde{J}}

\newcommand{\hc}{\hat{c}}

\newcommand{\hgamma}{\widehat{\gamma}}
\newcommand{\hdelta}{\widehat{\delta}}

\newcommand{\ket}[1]{{\left|#1\right\rangle}}

\newcommand{\Th}[2]{\Theta_{#1,#2}}
\renewcommand{\th}{{\theta}}

\newcommand{\ch}[2]{\mbox{ch}^{#1}_{#2}}

\newcommand{\tr}{\mbox{Tr}}
\renewcommand{\mod}{\mbox{mod}}

\newcommand{\nn}{\nonumber\\}

\newcommand{\NS}{\mbox{NS}}
\newcommand{\tNS}{\widetilde{\mbox{NS}}}
\newcommand{\R}{\mbox{R}}
\newcommand{\tR}{\widetilde{\mbox{R}}}
\newcommand{\sNS}{\msc{NS}}
\newcommand{\stNS}{\widetilde{\msc{NS}}}
\newcommand{\sR}{\msc{R}}
\newcommand{\stR}{\widetilde{\msc{R}}}

\newcommand{\any}{{}^{\forall}}
\newcommand{\ex}{{}^{\exists}}

\newcommand{\bigbox}[2]{{\scs{#1}} \hspace{0.3mm}  \raise-1mm\hbox{$ \underset{#2} {\scalebox{1.9}{\mbox{$\Box$}}} $}}

\newcommand{\eqn}[1]{(\ref{#1})}

\begin{document}

%%%%%%%%%%%%%%%%%%%%%%%%%%%%%%%%%%%%%%%%%%%%%%%%%%%%%%%%
%%%%%%%%%%%%%%%%%%%%%%%%%%%%%%%%%%%%%%%%%%%%%%%%%%%%%%%%
%%%%%%%%%%%%%%%%%%%%%%%%%%%%%%%%%%%%%%%%%%%%%%%%%%%%%%%%

%%% Title page %%%%%
\begin{titlepage}
 \
 \renewcommand{\thefootnote}{\fnsymbol{footnote}}
 \font\csc=cmcsc10 scaled\magstep1
 {\baselineskip=16pt
  \hfill
% \vbox{\hbox{January, 2020}}
}

 \baselineskip=20pt
\vskip 1cm
 
\begin{center}

{\bf \Large 

Non-SUSY Gepner Models

with Vanishing Cosmological Constant

} 

 \vskip 1.2cm

\noindent{ \large Koji Aoyama}\footnote{\sf ro0018hk@ed.ritsumei.ac.jp},

\vspace{2mm}

%\hspace{1cm}

\noindent{ \large Yuji Sugawara}\footnote{\sf ysugawa@se.ritsumei.ac.jp},
%%%

\medskip

 {\it Department of Physical Sciences, 
 College of Science and Engineering, \\ 
Ritsumeikan University,  
Shiga 525-8577, Japan}

\end{center}

\bigskip

\begin{abstract}

In this article we discuss a construction of non-SUSY type II string vacua 
with the vanishing cosmological constant at the one loop level based on the generic Gepner models for Calabi-Yau 3-folds. 
We make an orbifolding of the Gepner models by $\bz_2 \times \bz_4$, 
which  asymmetrically acts with some discrete torsions incorporated.   
We demonstrate that the obtained type II string vacua indeed lead to the vanishing cosmological constant at the one loop, 
whereas any space-time supercharges cannot be constructed as long as assuming the chiral forms 
 such as
`$
\dsp 
\cQ^{\al}_L \equiv \oint dz\, \cJ_L^{\al}(z)
$'.
We further discuss possible generalizations of the models described above.

\end{abstract}

\setcounter{footnote}{0}
\renewcommand{\thefootnote}{\arabic{footnote}}

\end{titlepage}

\baselineskip 18pt

\vskip2cm 
\newpage

%%%%%%%%%%%%%%%%%%%%%%%%%%%%%%%%%%%%%%%%
%%%%%%%%%%%%%%%%%%%%%%%%%%%%
%%%%%%%%%%%%%%%%%%%%%%%%%%%%%%%%%%%%%%%%
%%%%%%%%%%%%%%%%%%%%%%%%%%%%

\section{Introduction}

String theories on the {\em non-geometric\/} backgrounds would lead to  
interesting aspects not realized in the geometric ones.  
%%%%
It is worthy of special mention that the vanishing cosmological constant 
(at least at the level of one-loop) could be realized with no helps of unbroken SUSY 
in such non-geometric string vacua.
%%%
The studies on non-SUSY string vacua with vanishing cosmological constant 
have been initiated by \cite{Kachru1,Kachru2,Kachru3} 
based on some non-abelian orbifolds, followed by studies {\em e.g.} in 
\cite{Harvey,Shiu-Tye,Blumenhagen:1998uf,Angelantonj:1999gm,Antoniadis,Aoki:2003sy}. 
More recently, several non-SUSY vacua with this property have been  constructed  
as asymmetric orbifolds \cite{Narain:1986qm} 
by simpler cyclic groups 
in \cite{SSW,SWada}.

%%%%%%%%%%%%%%%%%%%%%%%%%%%%%%%%%%%%%%%%%%%%%%%%%%%%%%%%%%%%%%%%%%%%%%%%%%%%%%%%%%%%%%%%%%%%%%%%%%%%%%%

However, 
as far as we know, 
these attempts  
%to construct such non-SUSY string vacua 
have been limited to the toroidal models, 
which are realized as some asymmetric orbifolds of tori at  particular points of the moduli space. 
Therefore, it is  interesting to try to construct 
the {\em non-toroidal\/} string vacua possessing these properties.

Another motivation of this study is to search the non-SUSY heterotic string vacua with an {\em exactly} vanishing cosmological constant
at the one-loop. Notice that, in the known models of heterotic string vacua, one can only gain 
the small cosmological constant exponentially suppressed with respect to some moduli ({\em e.g} the radii of tori of compactifications) 
in the early study \cite{Harvey}, and also in 
closely related works given {\em e.g.} in \cite{Blaszczyk:2014qoa,Angelantonj:2014dia,Faraggi:2014eoa,
%%%%
%Dience,
Abel:2015oxa,
Kounnas:2015yrc,Abel:2017rch,
%%%
GrootNibbelink:2017luf
%%%
}. 
%%%

In this paper, we start with the Gepner constructions for Calabi-Yau 3-folds, which are generic enough, 
and make an attempt to construct the type II string vacua by considering some asymmetric orbifolds,   
in which we have a vanishing cosmological constant at the one loop while we cannot compose the space-time supercharges with a reasonable form. 
We believe this work to be the first attempt to construct the non-SUSY string vacua with the properties mentioned above based on the Gepner constructions. 
%%%%%%%%%%%%%%%%%%
We are still limited to the type II cases, but our approach could be extended to some heterotic string compactifications.  
We would like to report on the non-SUSY heterotic string vacua with the exactly vanishing cosmological constant as a future study.  
%%%%%%%%%%%%%%%%%%

~

%%%%%%%%%%%%%%%%%%%%%%%%%%%%%%%%%%%%%

This paper is organized as follows: 
We start with a very brief review of Gepner constructions \cite{Gepner} in section 2, 
mainly aiming at the preparations of notations.   
We also yield the definitions of orbifold actions utilized in our construction of 
string vacua. 
In section 3, we shall demonstrate our main results. Namely, we propose 
particular  
%non-trivial 
non-SUSY type II string vacua 
based on the asymmetric orbifolds of generic Gepner constructions of Calabi-Yau 3-folds, and show that the obtained vacua 
indeed realize vanishing cosmological constants at the one-loop, although any space-time supercharges with reasonable forms cannot be gained.
%due to the relevant orbifold actions. 
%%%%
We first consider the simplest case of $\bz_2 \times \bz_4$-orbifold, and then present
possible generalizations of it.  
%%%%
In section 4, 
we address the comparisons of the present models with the toroidal ones 
given in \cite{SSW,SWada} as discussions.

~

%%%%%%%%%%%%%%%%%%%%%%%%%%%%%%%%%%%%%%%%%%%%%%%%%%%%%%%%%%%%%%%%%%%%%
%%%%%%%%%%%%%%%%%%%%%%%%%%%%%%%%%%%%%%%%%%%%%%%%%%%%%%%%%%%%%%%%%%%%%
%%%%%%%%%%%%%%%%%%%%%%%%%%%%%%%%%%%%%%%%%%%%%%%%%%%%%%%%%%%%%%%%%%%%%

%%%%%%%%%%%%%%%%%%%%%%%%%%%%%%%%%%%%%%%%%%%%%%%%%%%%%%%
% Gepner model
%%%%%%%%%%%%%%%%%%%%%%%%%%%%%%%%%%%%%%%%%%%%%%%%%%%%%%%

\section{Preliminaries}
\label{sec:pre}

In this preliminary section we summarize the Gepner construction \cite{Gepner} and the orbifold actions that we will 
utilize in the next section.

~

%%%%%%%%%%%%%%%%%%%%%%%%%%%%%%%%%%%%%%%%%%%%%%%%%%%%%%%%%%
%%%%%%%%%%%%%%%%%%%%%%%%%%%%%%%%%%%%%%%%%%%%%%%%%%%%%%%%%%

\subsection{Gepner Models for $CY_3$}
\label{Gepner}

Let us consider the generic Gepner construction \cite{Gepner} for 
${\rm CY}_3$, that is, 
the superconformal system defined by
\begin{eqnarray}
 && \left\lb \cM_{k_1}\otimes \cdots  \otimes
\cM_{k_r}\right\rb\left|_{\bsz_N\msc{-orbifold}}
\right. , ~~~ \sum_{i=1}^r \frac{k_i}{k_i+2}=3,
\label{Gepner CY3}
\end{eqnarray}
where $\cM_k$ denotes the $\cN=2$ minimal model of level $k$
($\dsp \hat{c}\equiv \frac{c}{3}= \frac{k}{k+2}$), and we 
set
\begin{eqnarray}
 N := \mbox{L.C.M.} \, \{k_i+2~;~i=1,\ldots,r\}.
 \label{def N}
\end{eqnarray}
%%%%
%Through this paper we shall assume that $N$ is even. 
%%%%
%%%%%%%%%%%%%%%%%%%%%
In order to realize the modular invariance manifestly, 
we start with simple products of the characters of $\cN=2$ minimal model \cite{Dobrev,RY1} in the NS-sector, 
%%%%
denoted as `$\ch{(\sNS)}{\ell_i, m_i}(\tau,z)$' with the representation labels $(\ell_i, m_i)$, 
$\left(0\leq \ell_i \leq k_i, ~ m_i \in \bz_{2(k_i+2)}, ~ \ell_i+m_i \in 2\bz\right)$ ;  
%%%%
%%%%%%%%%%%%%%%%%%%%%%%%%%%%%%%%%%%%%%%
%%%%%%%%%%%%%%%%%%%%%%%%%%%%%%%%%%%%%%%
\begin{align}
& F^{(\sNS)}_I (\tau,z) := \prod_{i=1}^r \, \ch{(\sNS)}{\ell_i, m_i}(\tau,z),
\hspace{1cm} \left( I \equiv \left\{(\ell_i, m_i) \right\}
%, ~  ~ \ell_i+m_i \in 2\bz, \any i 
\right),
\label{FNS 0}
\end{align}
%%%%%%%%%%%%%%%%%%%%%%%%%%%%%%%%%%%%%%%
%%%%%%%%%%%%%%%%%%%%%%%%%%%%%%%%%%%%%%%
as the fundamental building blocks\footnote{We summarize the character formulas of 
$\ch{(\sigma)}{\ell, m}(\tau,z)$ with the spin structure $\sigma$
as well as the convention of theta functions in appendix A.
We set $q:= e^{2\pi i \tau}$, $y:= e^{2\pi i z}$ through this paper.},
%%%
and construct the ones for other spin structures by 
making the half spectral flows $\dsp z\, \mapsto \, z+ \frac{r}{2} \tau + \frac{s}{2}$
($r,s \in \bz_2$):
%%%%%%%%%%%%%%%%%%%%%%%%%%%%%%%%%%%%%%%
\begin{align}
F^{(\stNS)}_I (\tau,z) & := F^{(\sNS)}_I \left(\tau,z+\frac{1}{2} \right),
\label{FtNS 0}
\\
F^{(\sR)}_I (\tau,z) & := q^{\frac{\hc}{8}} y^{\frac{\hc}{2}} \, F^{(\sNS)}_I\left(\tau, z+ \frac{\tau}{2} \right),
\label{FR 0}
\\
F^{(\stR)}_I (\tau,z) & := q^{\frac{\hc}{8}} y^{\frac{\hc}{2}} \, F^{(\sNS)}_I\left(\tau, z+ \frac{\tau+1}{2} \right),
\label{FtR 0}
\end{align}
where we set $\hc=3$. 
%%%%%%%%%%%%
%%%%%%%%%%%%
We note that $\tNS$ ($\tR$) denotes 
the spin structure in which the world-sheet fermions in the NS (R)-sector take the periodic boundary condition 
along the temporal directions on the Euclidean torus. 
%%%%%%%%%%%%
%%%%%%%%%%%%
Notice also that the label $I \equiv \left\{(\ell_i, m_i)\right\}$ of the building blocks 
(and the spectral flow orbits introduced below) expresses the quantum numbers for the NS-sector 
{\em even for\/} 
$F^{(\sR)}_I$ and $F^{(\stR)}_I$. 
%In other words, $\ell_i+m_i \in 2\bz$, $\any i=1,\ldots, r$
%%%%%%
%%%%%%%%%%%%%%%%%%%%%%%%%%%%%%%%%%%%%%%%%%%
%We summarize the explicit character formulas as well as the conventions of theta functions in appendix A. 
%%%%%%%%%%%%%%%%%%%%%%%%%%%%%%%%%%%%%%%%%%%

Furthermore,  we have to make the chiral $\bz_N\times \bz_N$ orbifolding by 
$g_L \equiv e^{2\pi i J^{\msc{tot}}_0}$ and $g_R \equiv e^{2\pi i \tJ^{\msc{tot}}_0}$, where 
$J^{\msc{tot}}$ ($\tJ^{\msc{tot}}$) expresses the  total $\cN=2$ $U(1)$-current in the left (right) mover acting over 
$\dsp \otimes_i \cM_{k_i}$. 
Recall that the zero-mode $J_0^{\msc{tot}}$ takes the eigen-values in $\dsp \frac{1}{N} \bz$ for the NS sector.
% of $\dsp \otimes_i \, \cM_{k_i}$. 
The chiral $\bz_N$-orbifolding (in the left-mover) is represented in a way respecting the good modular 
properties by considering the `spectral flow orbits' \cite{EOTY} defined as follows:
%%%%%%%%%%%%%%%%%%%%%%%%%%%%%%%%%%%%%%%%%%%%%%%%%%%%%%%%%
% spectral flow orbit
%%%%%%%%%%%%%%%%%%%%%%%%%%%%%%%%%%%%%%%%%%%%%%%%%%%%%%%%%
\begin{align}
%F^{(\sNS)}_{I, (a,b)} (\tau,z) & := q^{\frac{\hc}{2}a^2} y^{\hc a}F^{(\sNS)}_I \left(\tau,z+a\tau+b\right),
\cF^{(\sNS)}_I (\tau,z) & := \frac{1}{N} \sum_{a,b\in \bz_N} \, q^{\frac{\hc}{2}a^2} y^{\hc a}F^{(\sNS)}_I \left(\tau,z+a\tau+b\right),
\label{cFNS}
\\
% F^{(\stNS)}_{I, (a,b)} (\tau,z) &:= F^{(\sNS)}_{I, (a,b)} \left(\tau,z+\frac{1}{2}\right)
\cF^{(\stNS)}_{I} (\tau,z) &:= \cF^{(\sNS)}_{I} \left(\tau,z+\frac{1}{2}\right)
 \nn
 & \equiv \frac{1}{N} \sum_{a,b\in \bz_N} \, (-1)^{\hc a} q^{\frac{\hc}{2}a^2} y^{\hc a}F^{(\stNS)}_I \left(\tau,z+a\tau+b\right),
\label{cFtNS}
\\
\cF^{(\sR)}_{I} (\tau,z) &:= q^{\frac{\hc}{8}} y^{\frac{\hc}{2}}\cF^{(\sNS)}_{I} \left(\tau,z+\frac{\tau}{2}\right)
 \nn
 & \equiv \frac{1}{N} \sum_{a,b\in \bz_N} \, (-1)^{\hc b} q^{\frac{\hc}{2}a^2} y^{\hc a}F^{(\sR)}_I \left(\tau,z+a\tau+b\right),
\label{cFR}
\\
\cF^{(\stR)}_{I} (\tau,z) &:= q^{\frac{\hc}{8}} y^{\frac{\hc}{2}}\cF^{(\sNS)}_{I} \left(\tau,z+\frac{\tau+1}{2}\right)
 \nn
 & \equiv \frac{1}{N} \sum_{a,b\in \bz_N} \,  (-1)^{\hc (a+b)} q^{\frac{\hc}{2}a^2} y^{\hc a}F^{(\stR)}_I \left(\tau,z+a\tau+b\right).
\label{cFtR}
\end{align}
%%%%%%%%%%%%%%%%%%%%%%%%%%
%where we set $\hc=3$ in these expressions. 
%%%%%%%%%%%%%%%%%%%%%%%%%%%%%%%%%%%%%%%%%%%%%%%%%%%%%%%%%%%
We also use the abbreviated notation; $\cF^{(\sigma)}_I(\tau) \equiv \cF^{(\sigma)}_I(\tau,0)$.
%%%%
See Appendix B for the explicit forms of $\cF^{(\sigma)}_I(\tau,z)$ written
in terms of the $\cN=2$ minimal characters. 
%%%%

The modular invariant partition function (for the transverse part)
that describes the SUSY vacuum $\br^{3,1} \times \mbox{CY}_3 $
is now written as 
%%%%%%%%%%%%%%%%%%%%%%%%%%%%%%%%%%%%%%%%%%%%%%%%
% SUSY Gepner
%%%%%%%%%%%%%%%%%%%%%%%%%%%%%%%%%%%%%%%%%%%%%%%%
\begin{align}
Z_{\msc{SUSY}}(\tau, \bar{\tau}) & = \left(\frac{1}{\sqrt{\tau_2} \left|\eta\right|^2}\right)^2 \cdot 
\frac{1}{4 N} \, \sum_{\sigma_L, \sigma_R}\, 
\ep(\sigma_L)\ep(\sigma_R)\left(\frac{\th_{[\sigma_L]}}{\eta}\right)
\overline{\left(\frac{\th_{[\sigma_R]}}{\eta}\right)}
\nn
& \hspace{2cm}
\times \sum_{I_L,I_R}\, N_{I_L,I_R} \cF^{(\sigma_L)}_{I_L}(\tau)
\overline{\cF^{(\sigma_R)}_{I_R}(\tau)},
\label{SUSY Gepner}
\end{align}
%%%%%%%%%%%%%%
where the summations of $\sigma_L$, $\sigma_R$ are taken over $\sigma_L, \sigma_R = \NS, \tNS, \R, (\tR)$
as usual. 
%%%%%%%%%%%%%%%%%%%%%%%%%%%%%%%%%%%%%%%%%%%%%%%%
%%%%%%%%%%%%%%%%%%%%%%%%%%%%%%%%%%%%%%%%%%%%%%%%
We assume the modular invariant coefficient $N_{I_L, I_R}$ to be diagonal through this paper:    
\begin{equation}
N_{I_L, I_R} \equiv \prod_{i=1}^r\, \frac{1}{2} \delta_{\ell_{i, L}, \ell_{i, R}} \delta_{m_{i,L}, m_{i, R}}, 
\hspace{1cm} \left(I_L \equiv \left\{ (\ell_{i, L}, m_{i, L} ) \right\}, ~~~ I_R \equiv \left\{ (\ell_{i, R}, m_{i, R} ) \right\} \right).
\label{mod inv N}
\end{equation}
%%%%%%%%%%%%%%%%%%%%%%%%%%%%%%%%%%%%%%%%%%%%%%%%%
Here we set $\ep(\NS)= - \ep(\tNS) = - \ep(\R) = 1$ and $\th_{[\sNS]} \equiv \th_3(\tau,0)$, $\th_{[\stNS]} \equiv \th_4(\tau,0)$, 
$\th_{[\sR]} \equiv \th_2(\tau,0)$, $\left(\th_{[\stR]} \equiv - i \th_1(\tau,0) \equiv 0 \right)$ to describe the free fermion 
contributions.

%%%%%%%%%%%%%%%%%%%%%%%%%%%%%%%%%%%%%%%%%%%%%%%%%%%

We shall assume 
%$N_1 \equiv k_1+2 \in 4\bz$ 
$k_1+2 \equiv 4K \in 4\bz_{>0}$
so as to make the $\bz_2 \times \bz_4$-orbifolding by $\gamma$, $\delta $ defined below well-defined.

~

%%%%%%%%%%%%%%%%%%%%%%%%%%%%%%%%%%%%%%%%%%%%%%%%%%
%%%%%%%%%%%%%%%%%%%%%%%%%%%%%%%%%%%%%%%%%%%%%%%%%%
%%%%%%%%%%%%%%%%%%%%%%%%%%%%%%%%%%%%%%%%%%%%%%%%%%

\subsection{Orbifold Actions}
\label{subsec:orbifold actions}

Let us clarify the orbifold actions that we will utilize in order to construct the string vacua.

~

\begin{description}
\item[(i) $\gamma_L \in \bz_2$ : ]

~

We introduce an involution  $\gamma_L $  
which only acts on the left-mover of $\cM_{k_1}$-sector as the sign factor $(-1)^{\ell_{1,L}}$ for 
the `$SU(2)_{k_1}$-quantum number' $\ell_{1,L}$ irrespective of the spin structures. 
Namely, $\gamma_L$ commutes with all the generators of superconformal algebra, and 
the primary states of $\cM_{k_1}$
should be transformed as 
%%%%
\begin{equation}
\gamma_L ~ : ~ \ket{\ell_{1,L}, m_{1,L}}^{(\sigma_L)} 
~ \longmapsto ~ (-1)^{\ell_{1,L}} \, \ket{\ell_{1,L}, m_{1,L}}^{(\sigma_L)}.
\label{def gamma}
\end{equation}
%irrespective of the spin structure $(\sigma_L, \sigma_R)$.
%%%%

The twisted sector by $\gamma_L$ is slightly non-trivial: the primary states are of the forms as 
$ \ket{k_1- \ell_{1,L} , m_{1, L}}^{(\sigma_L)} 
%\otimes \ket{\ell_1, m_{1, R}}^{(\sigma_R)}
, $
and $\gamma_L$ acts on them as the {\em different\/} sign factor $(-1)^{\ell_{1,L} +1}$.
%%%%
This is required by the modular invariance. 
%%%%
In fact, the modular invariant  of the $\bz_2$-orbifold of affine $SU(2)_{k_1}$-theory 
by $\gamma_L \equiv (-1)^{\ell_{1,L}}$ is found to be
%%%%%%%%%%%%%%%%%%%%%%%%%%%%%%
% SU(2) gamma-orbifold
%%%%%%%%%%%%%%%%%%%%%%%%%%%%%%
\begin{align}
\left. Z^{SU(2)_{k_1}}(\tau,\bar{\tau}) \right|_{\msc{$\gamma_L$-orb}} & 
= \sum_{\ell_1=0, \, \ell_1 \in 2\bz}^{k_1}\, \left|\chi^{SU(2)_{k_1}}_{\ell_1}(\tau)\right|^2
+ \sum_{\ell_1=1, \, \ell_1 \in 2\bz+1}^{k_1-1}\, \chi^{SU(2)_{k_1}}_{k_1- \ell_1}(\tau) \overline{\chi^{SU(2)_{k_1}}_{\ell_1}(\tau)},
\label{Z SU(2) gamma-orb}
\end{align}
which coincides with the $D_{\frac{k_1}{2}+2}$-type modular invariant for $k_1 \in 4\bz+2$ \cite{CIZ,Kato}.

To summarize, $\gamma_L$ should act on the character of $\cM_{k_1}$-sector 
as follows;
%%%%%%%%%%%%%%%%%%%%%%%%%%%%%%%%%%%%%%%%%%%%%%%%%%
\begin{align}
& \gamma_{L, (a, b)} \cdot \ch{(\sigma)}{\ell_1,m_{1}}(\tau,z)  :=
\left\{ 
\begin{array}{ll}
(-1)^{b\ell_1} \, \ch{(\sigma)}{\ell_1,m_{1}}(\tau,z),  & ~~ (a=0), \\
(-1)^{b(\ell_1+1)} \, \ch{(\sigma)}{k_1- \ell_1,m_{1}}(\tau,z), & ~~ (a=1),
\end{array}
\right.
\label{gamma ch}
\end{align}
where $(a,b) \in \bz_2\times \bz_2$ labels the spatial and temporal twisting by $\gamma_L$.
%%%%%%%%%%%%%%%%%%%%%%%%%%%%%%%%%%%%%%%%%%%%%%%%%%
It is obvious that $\gamma_L$ preserves all the space-time SUSY, since it does not affect 
the integral spectral flows defining the spectral flow orbits $\cF^{(\sigma)}_I$.

~

%%%%%%%%%%%%%%%%%%%%%%%%%%%
%%%%%%%%%%%%%%%%%%%%%%%%%%%

\item[(ii) $\delta_R \in \bz_4 $ : ]

~

Nextly, we introduce an order 4 chiral operator $\delta_R$
%$\delta_R^n$ ($n\in \bz_4$) 
which acts only on the right-mover of  $\cM_{k_1}$-sector as 
\begin{equation}
\delta_R := e^{2\pi i \frac{k_1+2}{4} J^{(1)}_{R, 0}}, 
\end{equation}
where $J^{(1)}_R$ denotes the right-moving $U(1)$-current 
of $\cN=2$ SCA in the $\cM_{k_1}$-sector.
%%%
$\delta_R$ obviously commutes with all the generators of 
SCA and induces a $\bz_4$-phase factor 
$e^{2\pi i \frac{m_{1,R}}{4}}$ 
($e^{2\pi i \left( \frac{k_1+2}{8} + \frac{m_{1,R}}{4}\right)}$)
for the primary states 
$
%\ket{\ell_{1,R}, m_{1, L}}^{(\sigma_L)} \otimes 
\ket{\ell_{1,R} , m_{1, R}}^{(\sNS)} 
$
(
$
%\ket{\ell_{1, R} , m_{1, L}}^{(\sigma_L)} \otimes 
\ket{\ell_{1, R} , m_{1, R}}^{(\sR)} 
$
).

%%%%%%%%%%%%%%%%%%%%%%%%%%%%%%%%%%%%%%%%%%%%%%%%%%%%%%%%

The orbifold by $\delta_R$ is well described in terms of the spectral flow 
$\bar{z} \, \mapsto\, \bar{z}+ \frac{k_1+2}{4}\left(\al \bar{\tau} + \beta\right)$ ($\al, \beta \in \bz_4$).
Indeed, the non-trivial part of the spatially and temporally twisted sector labeled by $(\al, \beta)$ is explicitly represented 
in terms of the $\cN=2$ minimal characters 
%in the right-mover of $\cM_{k_1}$-sector 
as  
%%%%%%%%
\begin{align}
& \delta_{R, (\al,\beta)} \cdot \overline{\ch{(\sigma)}{\ell_1,m_{1}}(\tau,z)}  := \overline{ 
q^{\frac{k_1(k_1+2)}{32}\al^2} y^{\frac{k_1}{4} \al} 
e^{2\pi i \frac{k_1(k_1+2)}{32}\al\beta}\, \ch{(\sigma)}{\ell_1,m_{1}}\left(\tau,z+\frac{k_1+2}{4} \left(\al \tau + \beta\right)\right)
},
\nn
& \hspace{10cm}
(\al,\beta)\in \bz_4 \times \bz_4,
\label{delta ch}
\end{align}
{\em irrespective of the spin structure $\sigma$.}
%%%
The phase factor $e^{2\pi i \frac{k_1(k_1+2)}{32}\al\beta}$ appearing in \eqn{delta ch} is again required 
by the modular invariance, and will play a crucial role in our arguments given in the next section. 
%%%%%%%%%%%%%%%
We note that the R.H.S of \eqn{delta ch} indeed has the expected periodicities under $\al\, \rightarrow \, \al+ 4 $,
$\beta\, \rightarrow \, \beta + 4 $.
%%%%%%%%%%%%%%%

As is easily confirmed, the $\delta_R$-orbifolding completely breaks the right-moving space-time SUSY.
%For instance, the spectral flow orbits in the $\delta_R^{\al}$-twisted sector 
%$\delta_{R, (\al, 0)} \cdot \overline{ \cF_{I_R}^{(\sNS)}(\tau)}$ 
%for $\al \in 2\bz+1$ contain primary states with half-integral $U(1)$-charges 
%which spoils the locality of all the space-time supercharges. 
For instance, the $\delta_R$-projection leaves the primary states with  
the quantum numbers $m_{1,R}$ of the {\em same\/} oddity both in the NS and R-sectors 
(in other words, $\ell_{1,R}$ with the opposite oddity). Thus, it is impossible to combine
them into a supermultiplet by the action of half spectral flows.

~

%%%%%%%%%%%%%%%%%%%%%%%%%%%%%%%%%%%%%%%%%%%%%%%%%%%%%%%%%%

\item[(iii) $(-1)^{F_L}$ : ]

~

$F_L$ denotes the `space-time fermion number' in the left-mover, that is, 
$(-1)^{F_L}$ acts as the sign flip of the left-moving Ramond sector, which would often appear in the literatures of  thermal superstring theory
(see {\em e.g.} \cite{AW}).
($(-1)^{F_R}$ is defined in the same way for the right mover.)
%%%
Denoting the spatial and temporal twisting by the operator $(-1)^{F_L}$ as $\left[(-1)^{F_L}\right]_{(a,b)}$, 
$\left( (a,b) \in \bz_2 \times \bz_2 \right)$, its action to the spectral flow 
orbit $\cF^{(\sigma)}_I(\tau)$ is summarized as follows
(for $\sigma = \NS, ~ \tNS, ~ \R$)\footnote
   {Though it is not necessary for our purpose, we also note $\ep(\tR; 0,1) = \ep(\tR; 1, 0) = \ep(\tR; 1,1) = -1$. 
(see {\em e.g.} \cite{AW}).
};
\begin{align}
& \left[(-1)^{F_L}\right]_{(a,b)} \cdot \cF^{(\sigma)}_{I_L}(\tau) = \ep(\sigma; a,b) \, \cF^{(\sigma)}_{I_L}(\tau),
\nn
& ~~~ \left\{
\begin{array}{l}
 \ep(\R; 0,1) = \ep(\tNS; 1, 0) = \ep(\NS; 1,1) = -1, 
\\
\ep(\sigma; a, b) =1 ~~ \mbox{otherwise.}  
\end{array}
\right.  
\label{-1^F_L}
\end{align}
It is clearly compatible with the modular covariance. 
%%%
Namely, $(a,b) \in \bz_2 \times \bz_2$ behaves as the suitable doublet 
of $SL(2,\bz)$ by modular transformations. 
%%%

\end{description}

~

%%%%%%%%%%%%%%%%%%%%%%%%%%%%%%%%%%%%%%%%%%%%%%%%%%%

%We note that the $\gamma$ and $\delta$-actions commute with each other. More precisely, 
%one can show the equality 
%\begin{align}
%& \gamma_{(a,b)} \, \delta_{(\al,\beta)} \cdot \ch{(\sigma)}{\ell_1,m_{1}}(\tau,z)
% = \delta_{(\al,\beta)} \, \gamma_{(a,b)} \cdot \ch{(\sigma)}{\ell_1,m_{1}}(\tau,z) ,
%\nn 
%& \hspace{5cm} (\any (a,b) \in \bz_2\times \bz_2, ~~ \any (\al,\beta) \in \bz_4 \times \bz_4),
%\end{align}
%when acting on the right-moving minimal characters $\ch{(\sigma)}{\ell,m}(\tau,z)$ in the $M_{k_1}$-sector. 

The $\gamma_L$ and $\delta_R$-orbifolding are obviously compatible 
and we can consider the $\bz_2 \times \bz_4$-orbifolds of the Gepner models generated by 
these operators. 
%%%
The corresponding  modular invariant partition function is 
written as 
%%%%%%%%%%%%%%%%%%%%%%%%%%%%%%%%%%%%%%%%%%%%%%%%
% chiral SUSY model
%%%%%%%%%%%%%%%%%%%%%%%%%%%%%%%%%%%%%%%%%%%%%%%%
\begin{align}
Z_{\msc{chiral SUSY}}(\tau, \bar{\tau}) & = \left(\frac{1}{\sqrt{\tau_2} \left|\eta\right|^2}\right)^2 \cdot 
\frac{1}{4 N} \, \sum_{\sigma_L, \sigma_R}\, 
\ep(\sigma_L)\ep(\sigma_R)\left(\frac{\th_{[\sigma_L]}}{\eta}\right)
\overline{\left(\frac{\th_{[\sigma_R]}}{\eta}\right)}
\nn
& \hspace{1cm}
\times \frac{1}{8} \sum_{a,b\in \bz_2}\, \sum_{\al,\beta \in \bz_4} \, \sum_{I_L,I_R}\, N_{I_L,I_R} \gamma_{L, (a,b)} \delta_{R, (\al,\beta)} \cdot 
\cF^{(\sigma_L)}_{I_L}(\tau)
\overline{\cF^{(\sigma_R)}_{I_R}(\tau)}.
\label{chiral SUSY model}
\end{align}
%%%%%%%%%%%%%%%%%%%%%%%%%%%%%%%%%%%%%%%%%%%%%%%%
%%%%%%%%%%%%%%%%%%%%%%%%%%%%%%%%%%%%%%%%%%%%%%%%
Since $\delta_R$ fully breaks the right-moving SUSY as mentioned above, 
we have an $\cN=1$ SUSY in 4-dim. which only comes from the left-mover in this string vacuum.

~

%%%%%%%%%%%%%%%%%%%%%%%%%%%%%%%%%%%%%%%%%%%%%%%%%%%
%%%%%%%%%%%%%%%%%%%%%%%%%%%%%%%%%%%%%%%%%%%%%%%%%%%
%%%%%%%%%%%%%%%%%%%%%%%%%%%%%%%%%%%%%%%%%%%%%%%%%%%

\section{Non-SUSY Models with Vanishing Cosmological Constant}

In this section we present our main results. 
We shall demonstrate the construction of our proposals of non-SUSY string vacua based on some 
asymmetric orbifolding of Gepner models. We then show that the constructed vacua  induce a vanishing cosmological constant 
(torus partition function), whereas any supercharges with the reasonable form cannot be made up.

~

%%%%%%%%%%%%%%%%%%%%%%%%%%%%%%%%%%%%%%%%%%%%%%%%%%%%%%%%%%%%%%%
%%%%%%%%%%%%%%%%%%%%%%%%%%%%%%%%%%%%%%%%%%%%%%%%%%%%%%%%%%%%%%%

\subsection{Construction of the Non-SUSY Models}

We  consider the $\bz_2 \times \bz_4$-orbifolding of the Gepner model 
%defined in 
\eqn{SUSY Gepner} 
by the operators 
%$\hgamma := (-1)^{F_R} \, \gamma_L $, $\hdelta := (-1)^{F_L}\, \delta_R $. 
\begin{equation}
\hgamma := (-1)^{F_R} \, \gamma_L , \hspace{1cm} \hdelta := (-1)^{F_L}\, \delta_R .
\label{def hgamma hdelta}
\end{equation}
%%%%
We shall also assume the discrete torsion \cite{Vafa:1986wx,Vafa:1994rv,Gaberdiel:2000fe} among the $\hgamma$ and $\hdelta$-actions, 
defined as 
%%%%%%%%%%%%%%%%%%%%%%%%%%%%%%%%%%%%%%%%%%%%%%%%%%%%%%%%%%%%%%%%%%%%
% discrete torison
%%%%%%%%%%%%%%%%%%%%%%%%%%%%%%%%%%%%%%%%%%%%%%%%%%%%%%%%%%%%%%%%%%%%
\begin{equation}
\xi \left( a, \al \, ; b ,\beta \right) := (-1)^{\left(K-1\right) \left(a\beta - b \al\right)},
\label{dtorsion}
\end{equation}
where the labels $a, b \in \bz_2$ and $\al, \beta \in \bz_4$ indicate respectively the $\hgamma$, $\hdelta$ twisted sectors
as presented in \eqn{chiral SUSY model}, for instance. 
($a$, $\al$ denote the spatial twistings, while $b$, $\beta$ do the temporal ones.)
%%%%%
Recall that we assumed $k_1+2 \equiv 4K \in 4 \bz_{>0}$.
%%%%%%%%%%%%%%%%%%%%%%%%%%%%%%%%%%%%%%%%%
The existence of this type torsion plays a crucial role to achieve the vanishing cosmological constant.

Now, we propose the string vacuum defined by the following 
modular invariant partition function: 
%%%%%%%%%%%%%%%%%%%%%%%%%%%%%%%%%%%%%%%%%%%%%%%%%
%%%%%%%%%%%%%%%%%%%%%%%%%%%%%%%%%%%%%%%%%%%%%%%%%
% non-SUSY Gepner
%%%%%%%%%%%%%%%%%%%%%%%%%%%%%%%%%%%%%%%%%%%%%%%%%
\begin{align}
%\hspace{-5mm}
Z_{\msc{non-SUSY}}(\tau, \bar{\tau}) & = \left(\frac{1}{\sqrt{\tau_2} \left|\eta\right|^2}\right)^2 \cdot 
\frac{1}{4 N} \, \sum_{\sigma_L, \sigma_R}\, 
\ep(\sigma_L)\ep(\sigma_R)\left(\frac{\th_{[\sigma_L]}}{\eta}\right)
\overline{\left(\frac{\th_{[\sigma_R]}}{\eta}\right)}
\nn
& 
%\hspace{5mm}
\times \frac{1}{8} \sum_{a,b\in \bz_2}\, \sum_{\al,\beta \in \bz_4} \, \sum_{I_L,I_R}\, N_{I_L,I_R} 
%%%
\xi \left(a,\al \, ; b, \beta \right)
%%%
\hgamma_{L,(a,b)} 
\hdelta_{R, (\al,\beta)} 
\cdot 
\cF^{(\sigma_L)}_{I_L}(\tau)
\overline{\cF^{(\sigma_R)}_{I_R}(\tau)}.
\label{non-SUSY Gepner}
\end{align}
%%%%%%%%%%%%%%%%%%%%%%%%%%%%%%%%%%%%%%%%%%%%%%%%
%%%%%%%%%%%%%%%%%%%%%%%%%%%%%%%%%%%%%%%%%%%%%%%%

Because of the definitions of orbifold actions 
\eqn{def hgamma hdelta}, 
it is obvious that the space-time SUSY is completely broken at least in the untwisted sector:
the right-moving SUSY is already broken in the `chiral SUSY model' \eqn{chiral SUSY model}, 
and furthermore, all the left-moving supercharges are  removed 
due to the inclusion of factor $(-1)^{F_L}$ in $\hdelta$.
We will later discuss more carefully why the space-time supercharges cannot be gained even if 
taking account of the degrees of freedom in the twisted sectors.

~

%%%%%%%%%%%%%%%%%%%%%%%%%%%%%%%%%%%%%%%%%%%%%%%%
%%%%%%%%%%%%%%%%%%%%%%%%%%%%%%%%%%%%%%%%%%%%%%%%

\subsection{Vanishing Cosmological Constant}

%\noindent
%{\bf $Z_{\msc{tot}} = 0$ : }

We next discuss that the torus partition function \eqn{non-SUSY Gepner} actually vanishes 
in spite of the lack of space-time SUSY.
To this aim we make use of the conventional notation: 
$
\dsp 
\bigbox{\hgamma^b\hdelta^\beta}{\hgamma^a \hdelta^{\al}}
$
in order to express the contribution to the torus partition function from  
the $\hgamma^a \hdelta^{\al}$, $\hgamma^b\hdelta^\beta$-twisted sectors along 
the spatial and temporal directions respectively. 
In other words, we can schematically write 
\begin{equation}
\bigbox{\hgamma^b\hdelta^\beta}{\hgamma^a \hdelta^{\al}} \equiv \xi(a, \al\,; b, \beta)\,
\tr_{\hgamma^a \hdelta^{\al}\msc{-twisted}} \, 
\left[ \hgamma^b \hdelta^\beta \, q^{L_0-\frac{\hc}{8}} \overline{q^{\tL_0-\frac{\hc}{8}}} \right],
\label{def bigbox}
\end{equation}
where $\xi(a, \al\,; b, \beta)$ denotes the discrete torsion mentioned above.

~

%%%%%%%%%%%%%%%%%%%%%%%%%%%%%%%%%%%%%%%%%%%%%%%%%%%%%%%

\begin{itemize}

\item
{\bf the sectors with even $\al$ :}

We first focus on the `even sectors' in which both of $\al,\beta$ are even. 
Since $\hdelta^2 = \delta_R^2$ holds, the left-moving SUSY is kept unbroken 
in these sectors, while the right-moving one is broken completely.   
We thus obtain
%%%%%%%%%%%%%%%%%%%%%%%%%%%%%%%%%%%%%%%%%%%%%%%%%
% even sector
%%%%%%%%%%%%%%%%%%%%%%%%%%%%%%%%%%%%%%%%%%%%%%%%%
\begin{align}
& \bigbox{\hgamma^b\hdelta^\beta}{\hgamma^a \hdelta^{\al}} =0, \hspace{1cm} \left(\any \al,\, 
\beta \in 2\bz, ~~ \any a,b \right).
\label{even sector}
\end{align}

However, $\hdelta^{\beta}$ includes $(-1)^{F_L}$ when $\beta$ is odd, 
and thus the left-moving SUSY is broken in this sector; 
\begin{align}
& \bigbox{\hgamma^b\hdelta^\beta}{\hgamma^a \hdelta^{\al}} \neq 0, ~~~
 \left(\any \al\in 2\bz, ~ \any \beta \in 2\bz+1 ~~ \any a,b \right).
\end{align}

One can also confirm that 
\begin{align}
 \sum_{b\in \bz_2}\, \sum_{\beta \in \bz_4}\,
\bigbox{\hgamma^b\hdelta^\beta}{\hgamma^a \hdelta^{\al}}  = 
 \sum_{b\in \bz_2}\, \sum_{\beta' \in \bz_2}\,
\bigbox{\hgamma^b\hdelta^{2 \beta' +1}}{\hgamma^a \hdelta^{\al}}
\neq 0, ~~~ 
\left(\any \al \in 2\bz, ~ \any a \right).
\end{align}
Here we made use of \eqn{even sector} for the first equality.
Moreover, the summation of $\beta' \in \bz_2$ leaves 
the states  in the $\cM_{k_1}$-sector
character $\ch{(\sigma)}{\ell_{1,R}, m_{1,R}}(\tau)$
with $m_{1,R} \in 2\bz$
for each spin structure. 
%%%
At this point, it is a slightly non-trivial fact that  
$\delta_R(\al,\beta)$ includes the extra phase factor 
\begin{equation}
e^{-2\pi i \frac{k_1(k_1+2)}{32}\al\beta} \equiv e^{-2\pi i \frac{K}{4} (2K-1) \al\beta},
\label{phase delta}
\end{equation}
which ensures the modular invariance, 
%%%%
as was mentioned around \eqn{delta ch}. 
%%%%
However, since we are assuming $\al \in 2\bz$ here, 
this phase factor does not affect the oddity of $m_{1,R}$ survived by 
the $\hdelta$-orbifolding.
The discrete torsion \eqn{dtorsion} does not alter it, too.

On the other hand, 
$\hgamma$ acts as 
\begin{equation}
\hgamma = \left\{
\begin{array}{ll}
(-1)^{\ell_{1,L}} & ~~ (*, \NS)\mbox{-sector} , ~(\any \al \in 2\bz, ~ \any a),
\\
(-1)^{\ell_{1,L}+1} & ~~ (*, \R)\mbox{-sector} , ~ (\any \al \in 2\bz, ~\any a), 
\end{array}
\right.
\label{hgamma action even}
\end{equation}
because of \eqn{gamma ch}, \eqn{-1^F_L}.
%and \eqn{mod inv N}.
Note that \eqn{dtorsion} does not affect it.

In this way, recalling that the modular invariant coefficients \eqn{mod inv N} are diagonal, 
we find that
the states with even $m_{1,R}$ finally survive after making the orbifold projections.

~

%%%%%%%%%%%%%%%%%%%%%%%%%%%%%%%%%%%%%%%%%%%%%%%%%%%%%%%%%%%%%%%%%%%%%%%%%

\item
{\bf the sectors with odd $\al$ :}

We next focus on the sectors with odd $\al$.
Since $\hdelta^{\al}$ includes $(-1)^{F_L}$, 
each contribution 
$\bigbox{\hgamma^b\hdelta^\beta}{\hgamma^a \hdelta^{\al}}$ 
does not vanish separately because of the lack of bose-fermi cancellation. 
However, we can show that these contributions totally vanish after summing over the temporal twisting $\beta$, $b$;
\begin{equation}
\sum_{b\in \bz_2}\, \sum_{\beta \in \bz_4} \, \bigbox{\hgamma^b\hdelta^\beta}{\hgamma^a \hdelta^{\al}}
=0, \hspace{1cm} (\any \al\in 2\bz+1, ~ \any a).
\label{total odd al sector}
\end{equation}

To be more precise, we can see 
\begin{equation}
\sum_{b\in \bz_2}\, \sum_{\beta' \in \bz_2} \, \left. \bigbox{\hgamma^b\hdelta^{2\beta'}}{\hgamma^a \hdelta^{\al}} \right|_{\msc{each spin structure}}
=0, \hspace{1cm}  (\any \al\in 2\bz+1, ~ \any a).
\label{total odd al sector 2}
\end{equation}
\eqn{total odd al sector}  obviously follows from the stronger one \eqn{total odd al sector 2} by taking the modular $T$-transformation of it.

To show \eqn{total odd al sector 2}, let us first recall the phase factor \eqn{phase delta}.
We thus find that 
$
\bigbox{\bm{1}}{\hgamma^{a} \hdelta^{\al}}+\bigbox{\hdelta^2}{\hgamma^{a} \hdelta^{\al}},
$
%$(\al \in 2\bz+1)$,
%%%%
%and 
%$
%\bigbox{\bm{1}}{\hgamma \hdelta^{\al}}+\bigbox{\hdelta^2}{\hgamma \hdelta^{\al}},
%$
%%%%%
$(\any a , ~  \any \al \in 2\bz+1)$
only contains states in  the characters $\ch{(\sigma)}{\ell_{1,R}, m_{1,R}}(\tau)$
with $m_{1,R} \equiv K ~ \mbox{mod} \, 2$
for each spin structure.

On the other hand, due to the discrete torsion \eqn{dtorsion}, 
$\hgamma$ effectively acts as the phase;
\begin{equation}
\hgamma = \left\{
\begin{array}{ll}
(-1)^{\ell_{1,L}+K-1} & ~~ (*, \NS)\mbox{-sector},  ~(\any \al\in 2\bz+1, ~ \any a),
\\
(-1)^{\ell_{1,L} +K} & ~~ (*, \R)\mbox{-sector},  ~ (\any \al\in 2\bz+1,~ \any a),
\end{array}
\right.
\label{hgamma action odd}
\end{equation}
in place of \eqn{hgamma action even}.
%in this odd $\al$-sector. 
%%%%
Therefore, after inserting 
the projection 
 $\frac{1+\hgamma}{2}$, 
%only leaves the even $\ell_{1,L}$-states for the $(\NS, *)$-sector, and the
%odd $\ell_{1,L}$-states for the $(\R, *)$-sector. 
no states survive in the $\hdelta^{\al}$-twisted sectors with odd $\al$, which proves \eqn{total odd al sector 2}.

\end{itemize}

~

%%%%%%%%%%%%%%%%%%%%%%%%%%%%%%%%%%%%%%%%%%%%%%%%%%%%%%%%
%%%%%%%%%%%%%%%%%%%%%%%%%%%%%%%%%%%%%%%%%%%%%%%%%%%%%%%%

In summary, we have found 
%%%%
\begin{itemize}
\item In each of the twisted sectors for $\hgamma^a \hdelta^{\al}$ with $a=0,1$, $\al \in 2\bz \cap \bz_4$, 
the space-time SUSY is completely broken and we obtain 
$$
Z_{\hgamma^a \hdelta^{\al}}(\tau,\bar{\tau}) \equiv \frac{1}{2\cdot 4} \, \sum_{b\in \bz_2}\, \sum_{\beta \in \bz_4}\, \bigbox{\hgamma^b\hdelta^\beta}{\hgamma^a \hdelta^{\al}} \neq 0.
$$

%%%%%%%%%%%%%%%%%%%%%%%%%%%%%%%%%%%%%%

\item In each of the twisted sectors for $\hgamma^a \hdelta^{\al}$ 
with $a=0,1$, $\al \in (2\bz+1) \cap \bz_4$, 
all the states are projected out by the orbifold projection.

\end{itemize}

%%%%%%%%%%%%%%%%%%%%%%%%%%%%%%%%%%%%%%%%%%%%%%%%%%%%%%%%
%%%%%%%%%%%%%%%%%%%%%%%%%%%%%%%%%%%%%%%%%%%%%%%%%%%%%%%%

~

Finally, we show that the total partition function 
\eqn{non-SUSY Gepner} indeed vanishes;
\begin{equation}
Z_{\msc{non-SUSY}}(\tau, \bar{\tau}) \equiv \sum_{a\in \bz_2}\, \sum_{\al \in 2\bz \cap \bz_4} \, Z_{\hgamma^a \hdelta^{\al}}(\tau, \bar{\tau}) = 0.
\label{Ztot}
\end{equation}

In fact, 
the modular S-transformation of \eqn{total odd al sector 2} yields
\begin{align}
& \sum_{a\in \bz_2}\, \sum_{\al \in \bz_2} \, \bigbox{\hdelta^{\beta}}{\hgamma^a \hdelta^{2 \al}} =
\sum_{a\in \bz_2}\, \sum_{\al \in \bz_2} \, \bigbox{\hgamma \hdelta^{\beta}}{\hgamma^a \hdelta^{2\al}} =0, ~~~
 \left(\any \beta \in 2\bz+1 \right).
\label{total odd beta sector}
\end{align}
Combining this result with \eqn{even sector}
as well as \eqn{total odd al sector 2}, 
we immediately obtain the desired fact \eqn{Ztot}.

~

%%%%%%%%%%%%%%%%%%%%%%%%%%%%%%%%%%%%%%%%%%%%%%%%%%%%%%%%%%
%%%%%%%%%%%%%%%%%%%%%%%%%%%%%%%%%%%%%%%%%%%%%%%%%%%%%%%%%%

A few comments are in order.

~

%%%%%%%%%%%%%%%%%%%%%%%%%%%%%%
% absence of tachyons
%%%%%%%%%%%%%%%%%%%%%%%%%%%%%%

\noindent
{\bf 1. } 
We note that the GSO-projection operator for the right-moving NS-sector 
acts with the opposite sign in the $\hgamma$- and $\hdelta^2$-twisted sectors, 
which would potentially lead to a tachyonic instability. 
However, the left-movers in these twisted sectors are correctly GSO-projected in each spin structure, 
and thus no tachyons appear after the level matching condition is imposed.

~

%%%%%%%%%%%%%%%%%%%%%%%%%%%%%%%%%
% no chiral supercharges
%%%%%%%%%%%%%%%%%%%%%%%%%%%%%%%%%

\noindent
{\bf 2. } 
In the string vacuum constructed above, the bose-fermi cancellation occurs in the left-mover, 
{\em after summing up over all the twisted sectors;}
\begin{equation}
\sum_{a, \al}\, \left[Z_{a,\al\, (\sNS, *)} (\tau,\bar{\tau}) 
+ Z_{a, \al\, (\sR, *)}(\tau,\bar{\tau})\right] =0,
\label{left cancellation}
\end{equation}
while it does not in the right-mover. 
%%%
One might thus wonder if the space-time SUSY would eventually survive in the left-mover. 
%%%%%%%%%%
%%%%%%%%%%

We here clarify why we nevertheless claim that  the space-time supercharges should be absent in our model. 
We first assume the chiral forms 
`$
\dsp 
\cQ^{\al}_L \equiv \oint dz\, \cJ_L^{\al}(z)
$'
with some holomorphic currents $\cJ_L^{\al}(z)$ 
for the supercharges to be searched. 
We believe it to be a fairly reasonable assumption, 
because it physically means that the supercharges in question should be conserved quantities on the world-sheet. 
%%%%%%%%%
Now,
it is obvious by our orbifold construction that the space-time supercharges, if any, cannot originate from the untwisted sector. 
In other words it should belong to the twisted sector. 
Observing the aspect of bose-femi cancellation mentioned above, the expected supercharges
have to be made up of
the operators that intertwine the untwisted sector with 
 the $\hgamma \hdelta^2$-twisted sector, if it would exist anyway.  
However, it is not possible under the above assumption, 
%as long as assuming the chiral form
% such as
%`$
%\dsp 
%\cQ^{\al}_L \equiv \oint dz\, \cJ_L^{\al}(z)
%$'
%with some holomorphic currents $\cJ_L^{\al}(z)$ 
%for the expected supercharge, 
%which we believe to be a fairly reasonable assumption,  
since the twist operator 
$\hgamma \hdelta^2\equiv \gamma_L \otimes \delta_R^2 (-1)^{F_R}$ {\em non-chirally\/} acts on 
the Hilbert space at hand,
%and thus alters 
changing 
the boundary conditions of fields in {\em both\/} of the left and right-movers. 
%%%%%%%%
%$$
%Q^a = \oint dz\, S^a(z)
%$$
%%%%%%%%
%We believe that the chiral form of supercharges is a fairly reasonable assumption, 
%because it physically means the conservation of supercharges to be searched on the world-sheet. 

~

%%%%%%%%%%%%%%%%%%%%%%%%%%%%%%%%%%%%%%%%%%%%%%%%%%%%%%%%%%%%%%
% unitarity
%%%%%%%%%%%%%%%%%%%%%%%%%%%%%%%%%%%%%%%%%%%%%%%%%%%%%%%%%%%%%%

\noindent
{\bf 3. } 
The unitarity of conformal system constructed above is readily confirmed, although 
it would often be subtle in models of asymmetric orbifolds, especially for the twisted sectors. 
Namely, the partition sum for  the each twisted sector $Z_{a,\al}(\tau,\bar{\tau})$ has a $q$-expansion with 
coefficients of positive integers. This fact is obvious because all the orbifolds actions, which are summarized in \ref{subsec:orbifold actions},
are manifestly compatible with unitarity.

~

%%%%%%%%%%%%%%%%%%%%%%%%%%%%%%%%%%%%%%%%%%%%%%%%%%%%%%%%%%%%%%%
%%%%%%%%%%%%%%%%%%%%%%%%%%%%%%%%%%%%%%%%%%%%%%%%%%%%%%%%%%%%%%%
%%%%%%%%%%%%%%%%%%%%%%%%%%%%%%%%%%%%%%%%%%%%%%%%%%%%%%%%%%%%%%%

\subsection{Generalization of the Non-SUSY Models}
\label{generalization}

Here we shall present a generalization of the non-SUSY Gepner models constructed above. 
We again start with the Gepner models for $\mbox{CY}_3$ 
in which we have $N \left(\equiv \mbox{L.C.M}\, \{ k_i+2 \}\right) \in 4 \bz$.
%%%%
We would like to construct the orbifolds in the manner similar to the previous ones, 
but in which the orbifold operators  (denoted as `$\hgamma$', `$\hdelta$' again)
act on the {\em multiple\/} factors of the $\cN=2$ minimal models.

%%%%%%%%%%%%%%%%%%%%%%%%%%%%%%%%%%%%%%%%%%%%%%%%%%%%%%%%%%%%%%%%%%

Let us fix a subsystem of the minimal models $\otimes_{i\in S}\, \cM_{k_i} $, 
$S \subset \{ 1, 2, \ldots, r \}$, on which the orbifold operators non-trivially 
act. We assume 
\begin{equation}
N' \equiv \mbox{L.C.M.}\, \left\{k_i+2 \, :\, i \in S \right\} \in 4 \bz.
\end{equation}
%and set $N' = 4K'$, $K' \in \bz$.
It is obvious that the total central charge of the subsystem $S$ is written in the form;
\begin{equation}
\hc_S \left(\equiv \sum_{i \in S}\, \frac{k_i}{k_i+2}\right) = \frac{2 M}{N'}, \hspace{1cm} 
\ex M \in \bz.
\label{hc S}
\end{equation}
%%%%%
We also fix a positive integer $L$ dividing $\dsp \frac{N'}{4}$
%$K'$, 
and also define $S_1 \subset S$ by 
\begin{equation}
S_1 := \left\{ i \in S \, :\, \frac{N'}{k_i+2} \in 2\bz+1 \right\}.
\label{def S_1}
\end{equation} 
Note that $S_1 \neq \phi$, since $N'$ is the L.C.M. of $\{ k_i+2\}_{i\in S}$. 

%%%%%%%%%%%%%%%%%%%%%%%%%%%%%%%%%%%%%%%%%%%%%%%%%%%%%%%

%%%%%%%%%%%%%%%%%%%%%%%%%%%%%%%%%%%%%%%%%%%%%%%%%%%%%%%

~

%%%%%%%%%%%%%%%%%%%%%%%%%%%%%%%%%%%%%%%%%%%

Under the preparations given above,
we define the orbifold actions as well as the discrete torsion 
that generalize those given in the previous subsection.
%%%
Note that the previous one  just corresponds to the case 
$S = S_1 = \left\{ i=1 \right\}$, $N' = k_1+2 = 4K 
%\left(\in 4\bz\right)
$, 
$\dsp L= \frac{k_1+2}{4} =K $ and 
$\dsp M = \frac{k_1}{2} = 2K-1
%\left(\in 2\bz+1\right)
$;

%%%%%%%%%%%%%%%%%%%%%%%%%%%%%%%%%%%%%%

\begin{description}
\item[(i) $\hgamma \in \bz_2$ : ]
~

We define
\begin{equation}
\gamma_L := \prod_{i\in S_1}\, (-1)^{\ell_{i, L}}.
%\hspace{1cm} \hgamma := (-1)^{F_R} \gamma_L.
\label{def hgamma S}
\end{equation}
Namely, $\gamma_L$ acts on the left-moving characters of each minimal model $\cM_{k_i}$, $\any i \in S_1$
as the $\bz_2$-twisting \eqn{gamma ch}.
%%%%%%%%%%%
We then set 
\begin{equation}
\hgamma := \left\{
\begin{array}{ll}
 (-1)^{F_R} \gamma_L & ~~ (\# S_1 \in 2\bz+1), \\
\gamma_L & ~~ (\# S_1 \in 2\bz) .
\end{array}
\right.
\label{hgamma general}
\end{equation}

~

%%%%%%%%%%%%%%%%%%%%%%%%%%%%%%%%%%%%%%%%%%%%%%%%%%%%%%

\item[(ii) $\hdelta \in \bz_{N'/L}$ : ]
~

We define 
\begin{equation}
\delta_R := e^{2\pi i L \sum_{i \in S}\, J_{R, 0}^{(i)}}, \hspace{1cm} \hdelta := (-1)^{F_L} \delta_R,
\label{def hdelta S}
\end{equation}
where $J_R^{(i)}$ is the right-moving $U(1)$-current in $\cM_{k_i}$, $\any i\in S$. 
In other words, $\delta_R$ acts on the right-moving characters of $\cM_{k_i}$, $\any i\in S$ as 
the integral spectral flow $\dsp \bar{z}\, \mapsto \, \bar{z} + L (\al \bar{\tau} + \beta)$;
\begin{align}
& \delta_{R, (\al,\beta)} \cdot \overline{\ch{(\sigma)}{\ell_i ,m_{i}}(\tau,z)}  := \overline{ 
q^{\frac{k_i}{2(k_i+2)}L^2 \al^2} y^{\frac{k_i}{k_i+2} L \al} 
e^{2\pi i \frac{k_i}{2(k_i+2)} L^2 \al\beta}\,
\ch{(\sigma)}{\ell_i,m_i}\left(\tau,z+L \left(\al \tau + \beta\right)\right)
},
\nn
& \hspace{10cm}
(\al,\beta)\in \bz_{N'/L} \times \bz_{N'/L} ,
\label{delta ch S}
\end{align}
as in \eqn{delta ch}.

~

%%%%%%%%%%%%%%%%%%%%%%%%%%%%%%%%%%%%%%%%%%%%%%%%%%%%%%%%

\item[(iii) discrete torsion : ]
~

We also introduce the discrete torsion with respect to the $\hgamma$ and $\hdelta$-actions as in 
\eqn{dtorsion};
%%%%%%%%%%%%%%%%%%%%%%%%%%%%%%%%%%%%%%%%%%%%%%%%%%%%%%%%%%%%%%%%%%%%
% discrete torison
%%%%%%%%%%%%%%%%%%%%%%%%%%%%%%%%%%%%%%%%%%%%%%%%%%%%%%%%%%%%%%%%%%%%
\begin{equation}
\xi \left( a, \al \, ; b ,\beta \right) := (-1)^{\left(LM-1\right) \left(a\beta - b \al\right)},
\hspace{1cm} \left(a, b \in \bz_2, ~ \al, \beta \in \bz_{N'/L} \right).
\label{dtorsion S}
\end{equation}

\end{description}
%%%%

%~

%Note that the previous one  just corresponds to the case 
%$S = S_1 = \left\{ i=1 \right\}$, $N' = k_1+2 = 4K 
%%\left(\in 4\bz\right)
%$, 
%$\dsp L= \frac{k_1+2}{4} =K $ and 
%$\dsp M = \frac{k_1}{2} = 2K-1
%%\left(\in 2\bz+1\right)
%$

~

%%%%%%%%%%%%%%%%%%%%%%%%%%%%%%%%%%%%%%%%%%%%%%%%%%%%%%%%%
%%%%%%%%%%%%%%%%%%%%%%%%%%%%%%%%%%%%%%%%%%%%%%%%%%%%%%%%%

We can now define the $\bz_2 \times \bz_{N'/L}$-orbifold of the Gepner model at hand,  
%$\bigotimes_i \cM_{k_i}$
which describes a non-SUSY string vacuum. 
We can further show that the torus partition function vanishes;
\begin{equation}
Z_{\msc{non-SUSY}}(\tau, \bar{\tau}) \equiv \sum_{a,b\in \bz_2}\, 
\sum_{\al,\beta \in \bz_{N'/L}}\,
 \bigbox{\hgamma^b \hdelta^{\beta}}{\hgamma^a \hdelta^{\al}} =0,
\label{total Z S}
\end{equation}
while 
\begin{align}
%Z_{\hgamma^a \hdelta^{\al}}(\tau,\bar{\tau}) 
Z_{a,\al}(\tau,\bar{\tau})
\equiv 
\sum_{b\in \bz_2}\, 
\sum_{\beta \in \bz_{N'/L}}\,
 \bigbox{\hgamma^b \hdelta^{\beta}}{\hgamma^a \hdelta^{\al}} \neq 0,
%\hspace{1cm} \left(\any a \in \bz_2, ~ \any \al \in \bz_{\frac{N'}{L}} \cap 2\bz \right).
\label{total Z twisted sector S}
\end{align}
for each twisted sector with $a\in \bz_2$ and $\al \in \bz_{N'/L} \cap 2\bz $.

%%%%%%%%%%%%%%%%%%%%%%%%%%%%%%%%%%%%%%%%%%%%%%%%%%%%%%%%%%%%%%%%%%%%%

To show it, the next fact plays a crucial role; 
\begin{equation}
\sum_{b\in \bz_2}\, \sum_{\beta \in \bz_{N'/L} \cap 2\bz} \, \left. \bigbox{\hgamma^b\hdelta^{\beta}}{\hgamma^a \hdelta^{\al}} \right|_{\msc{each spin structure}}
=0, 
%\hspace{1cm}  (\any \al\,:\,\mbox{odd}, ~ \any a).
\label{total odd al sector 2 S}
\end{equation}
for any `odd sectors' with $\any \al \in \bz_{N'/L} \cap \left(2\bz+1\right)$ and 
$\any a \in \bz_2$, which is the analogue of \eqn{total odd al sector 2}.
%%%%%%%%%%%%%%%%%%%%

In fact, the summation over $\beta \in \bz_{N'/L} \cap 2\bz$ imposes the constraint 
\begin{align}
N' \left[\sum_{i\in S}\, \frac{m_{i, R}-2 n_R}{k_i+2} + \frac{\hc_S}{2}L \al\right]
& \equiv \sum_{i\in S}\, d_i (m_{i, R}-2n_R) + LM \al \in \frac{N'}{2L}\bz 
\hspace{1cm} \left(d_i \equiv \frac{N'}{k_i+2} \right),
\label{eval beta proj S}
\end{align}
on the characters $\ch{(\sigma)}{\ell_{i,R}, m_{i,R}-2n_R}(\tau,z)$ ($i \in S$).
%%%%
($n_R \in \bz_N$ denotes the spectral flow parameter appearing in the right-moving 
orbit $\cF^{(\sigma)}_{I_R}(\tau)$.)
%%%%
Recalling our assumption that $d_i \in 2\bz+1$ iff $i \in S_1$ and 
$\dsp \frac{N'}{2L}
%=\frac{2 K'}{L} 
\in 2\bz$, 
this relation implies 
\begin{equation}
\sum_{i\in S_1}\, m_{i,R} \equiv LM ~ \mod\, 2,
\label{eval beta proj S 2}
\end{equation}
in each spin structure.
In other words, 
%%%%%
\begin{itemize}
\item for $\# S_1 \in 2\bz+1$, we obtain
\begin{equation}
\sum_{i\in S_1}\, \ell_{i,R} \equiv 
\left\{
\begin{array}{ll}
LM ~ \mod\, 2, & ~~ \mbox{$(*, \NS)$-sector},
\\
LM+1 ~ \mod\, 2, & ~~ \mbox{$(*, \R)$-sector},
\end{array}
\right.
\label{eval beta proj S 3}
\end{equation}

\item for $\# S_1 \in 2\bz$, we obtain
\begin{equation}
\sum_{i\in S_1}\, \ell_{i,R} \equiv LM ~ \mod\, 2,
\label{eval beta proj S 4}
\end{equation}
irrespective of the spin structure.

\end{itemize}

%%%%%

On the other hand, taking account of the discrete torsion \eqn{dtorsion S}, we find that 
$\hgamma$ effectively acts as   
%%%%%
\begin{itemize}
\item for $\# S_1 \in 2\bz+1$,
\begin{equation}
\hgamma = \left\{
\begin{array}{ll}
(-1)^{\sum_{i\in S_1} \ell_{i,L} +LM-1} & ~~ (*, \NS)\mbox{-sector},  ~(\any \al\in 2\bz+1, ~ \any a),
\\
(-1)^{\sum_{i\in S_1} \ell_{i,L} +LM} & ~~ (*, \R)\mbox{-sector},  ~ (\any \al\in 2\bz+1,~ \any a),
\end{array}
\right.
\label{hgamma action odd S}
\end{equation}
as in \eqn{hgamma action odd}, while
%%%%%%%%%%%%%%%%%%%

\item for $\# S_1 \in 2\bz$,
\begin{equation}
\hgamma = (-1)^{\sum_{i\in S_1} \ell_{i,L} +LM-1}, ~~~ ~(\any \al\in 2\bz+1, ~ \any a),
\label{hgamma action odd S 2}
\end{equation}
irrespective of the spin structure. 

\end{itemize}

By comparing 
\eqn{eval beta proj S 3}, \eqn{eval beta proj S 4} with \eqn{hgamma action odd S},
\eqn{hgamma action odd S 2}
one can show that the contributions in question vanish separately in each spin structure  
after making 
$\hgamma$-projection, proving 
the fact \eqn{total odd al sector 2 S}. 
%%%
We thus obtain the desired result \eqn{total Z S} according to the same argument 
as given in the previous subsection.

%%%%%%%%%%%%%%%%%%%%%%%%%%%%%%%%%%%%%%%%%%%%%%%%%%%%%%%%%%%%%%%%%
%%%%%%%%%%%%%%%%%%%%%%%%%%%%%%%%%%%%%%%%%%%%%%%%%%%%%%%%%%%%%%%%%
%%%%%%%%%%%%%%%%%%%%%%%%%%%%%%%%%%%%%%%%%%%%%%%%%%%%%%%%%%%%%%%%%

~

\section{Discussions}

In this paper, we have studied some asymmetric orbifolds of the Gepner models for Calabi-Yau 3-folds, 
aiming at the construction of non-SUSY type II string vacua with the vanishing cosmological 
constant at one-loop.

%%%%%%%%%%%%%%%%%%%%%%%%%%%%%%%%%%%%%%%%%%

We would like to compare several aspects of the present model with those of the ones 
given in \cite{SSW,SWada},
which are constructed as asymmetric orbifolds of tori. 
%%%%%%%%%%%%%%

In the ones adopted in \cite{SSW,SWada}, 
the asymmetric orbifold actions 
that generate the `SUSY-breaking factors' $(-1)^{F_L}$, $(-1)^{F_R}$ have been combined 
with the translation along some direction in the compactification space. 
This would be an analogue of Scherk-Schwarz type compactification \cite{SS}. 
It is a characteristic feature of this model  that the bose-fermi cancellation occurs at 
the {\em each} sector corresponding to \eqn{def bigbox} in this paper. 
Indeed, the {\em left-moving\/} bose-fermi 
cancellation occurs in the  sectors with even winding numbers along the `Scherk-Schwarz circle',
whereas we have the  {\em right-moving\/} bose-fermi cancellation 
in the odd winding sectors. 
This aspect prevents us from constructing any supercharges defined over the total Hilbert space.

In the present model
%on the other hand, 
the orbifold actions $\hgamma$, $\hdelta$ likewise include 
%the factors 
$(-1)^{F_L}$, $(-1)^{F_R}$. 
On the other hand, we did not assume the Scherk-Schwarz circle in any direction of compactification.  
Indeed, we started with a Gepner model for $\mbox{CY}_3$ 
and the Scherk-Schwarz type compactification seems to be hard to make, since the translational invariance 
is generically broken.

Another crucial difference is that  we do not have the bose-fermi cancellation 
in {\em each\/} twisted sector in the present model. 
Namely
$
\dsp 
Z_{a, \al} \equiv  \sum_{b, \beta} \, \bigbox{\hgamma^b \hdelta^{\beta}}{\hgamma^a \hdelta^{\al}} 
$ 
for fixed $a$, $\al$
%as well as each $\dsp \bigbox{\hgamma^b \hdelta^{\beta}}{\hgamma^a \hdelta^{\al}} $
does not necessarily vanish.
Nevertheless, the total partition function vanishes {\em after summing up over all the twisted sectors}:
$$
Z \equiv \sum_{a, \al} Z_{a, \al} \equiv  \sum_{a,\al} \,\sum_{b, \beta} \, \bigbox{\hgamma^b \hdelta^{\beta}}{\hgamma^a \hdelta^{\al}} =0.
$$
%%%
This feature is in a sharp contrast with the previous ones. 
%%%
To be more specific, we have the bose-fermi cancellation {\em only in the left-mover,}
as was noted around \eqn{left cancellation}. 
Nonetheless we cannot gain the left-moving supercharges because of the non-chirality of 
the orbifold actions $\hgamma$, $\hdelta$.

We would also like to point out that the unitarity is rather simple to confirm in the present model, 
though it was  non-trivial whether the torus partition functions are $q$-expanded in the way 
compatible with the unitarity in the models given in \cite{SSW,SWada}.

%%%%%%%%%%%%%%%%%%%%%%%%
% extension to hetero
%%%%%%%%%%%%%%%%%%%%%%%%

Since the right-mover does not play any role in achieving the vanishing 
cosmological constant,
the present construction could be applicable to the heterotic string vacua, too, 
whereas it was difficult for the previous ones in \cite{SSW,SWada}, in which 
both of the left and right-moving bose-fermi cancellations are necessary for realizing 
the desired non-SUSY vacua. 
We would like to make the detailed studies of extensions to the  
heterotic string vacua in a future work.

~

%%%%%%%%%%%%%%%%%%%%%%%%%%%%%%%%%%%%%%%%%%%%%%%%%%%%%%%%%%%%%%%%
%%%%%%%%%%%%%%%%%%%%%%%%%%%%%%%%%%%%%%%%%%%%%%%%%%%%%%%%%%%%%%%%
%%%%%%%%%%%%%%%%%%%%%%%%%%%%%%%%%%%%%%%%%%%%%%%%%%%%%%%%%%%%%%%%
%%%%%%%%%%%%%%%%%%%%%%%%%%%%%%%%%%%%%%%%%%%%%%%%%%%%%%%%%%%%%%%%

%\section*{Acknowledgments}

%\newpage

%%%%%%%%%%%%%%%%%%%%%%%%%%%%%%%%%%%%%%%%%%%%%%%%%%%%%%%%%%%
%%%%%%%%%%%%%%%%%%%%%%%%%%%%%%%%%%%%%%%%%%%%%%%%%%%%%%%%%%%
%%%%%%%%%%%%%%%%%%%%%%%%%%%%%%%%%%%%%%%%%%%%%%%%%%%%%%%%%%%
%%%%%%%%%%%%%%%%%%%%%%%%%%%%%%%%%%%%%%%%%%%%%%%%%%%%%%%%%%%

\appendix

\section*{Appendix A: ~ Summary of  Conventions}

\setcounter{equation}{0}
\def\theequation{A.\arabic{equation}}

~

We summarize the notations and conventions adopted in this paper. 
We set $q \equiv e^{2\pi i \tau}$, $y \equiv e^{2\pi i z}$.

~

%%%%%%%%%%%%%%%%%%%%%%%%%%%%%%%%%%%%%%%%%%
%%%%%%%%%%%%%%%%%%%%%%
\noindent
{\bf 1. Theta Functions} 
%%%%%%%%%%%%%%%%%%%%%%%%%%%%
%
%%%%%%%%%%%%%%%%%%%%%%
 \begin{align}
 & \dsp \th_1(\tau,z):=i\sum_{n=-\infty}^{\infty}(-1)^n q^{(n-1/2)^2/2} y^{n-1/2}
  \equiv  2 \sin(\pi z)q^{1/8}\prod_{m=1}^{\infty}
    (1-q^m)(1-yq^m)(1-y^{-1}q^m), \nn [-10pt]
   & \\[-5pt]
 & \dsp \th_2(\tau,z):=\sum_{n=-\infty}^{\infty} q^{(n-1/2)^2/2} y^{n-1/2}
  \equiv 2 \cos(\pi z)q^{1/8}\prod_{m=1}^{\infty}
    (1-q^m)(1+yq^m)(1+y^{-1}q^m), \\
 & \dsp \th_3(\tau,z):=\sum_{n=-\infty}^{\infty} q^{n^2/2} y^{n}
  \equiv \prod_{m=1}^{\infty}
    (1-q^m)(1+yq^{m-1/2})(1+y^{-1}q^{m-1/2}),  
\\
 &  \dsp \th_4(\tau,z):=\sum_{n=-\infty}^{\infty}(-1)^n q^{n^2/2} y^{n}
  \equiv \prod_{m=1}^{\infty}
    (1-q^m)(1-yq^{m-1/2})(1-y^{-1}q^{m-1/2}) . 
\\
& \Th{m}{k}(\tau,z):=\sum_{n=-\infty}^{\infty}
 q^{k(n+\frac{m}{2k})^2}y^{k(n+\frac{m}{2k})} ,
\\
&
\eta(\tau) := q^{1/24}\prod_{n=1}^{\infty}(1-q^n).
 \end{align}
 Here, we have set $q:= e^{2\pi i \tau}$, $y:=e^{2\pi i z}$  
 ($\any \tau \in \bh^+$, $\any z \in \bc$),
 and used abbreviations, $\th_i (\tau) \equiv \th_i(\tau, 0)$
 ($\th_1(\tau)\equiv 0$), 
$\Th{m}{k}(\tau) \equiv \Th{m}{k}(\tau,0)$.
%
%%%%%%%%%%%%%%%%%%%%%%%%%%%%%%%%%%%%%%%%%%%%%%%%%%%%%%%%%%%%%%%%%
%%%%%%%%%%%%%%%%%%%%%%%%%%%%%%%%%%%%%%%%%%%%%%%%%%%%%%%%%%%%%%%%%

~

%%%%%%%%%%%%%%%%%%%%%%%%%%%%%%%%%%%%%%%%%%%%%%%%%%%%%%%%%%%%%%%%%
%%%%%%%%%%%%%%%%%%%%%%%%%%%%%%%%%%%%%%%%%%%%%%%%%%%%%%%%%%%%%%%%%

\noindent
{\bf 2. Character Formulas for $\cN=2$ Minimal Model}

The character formulas of 
the level $k$ $\cN=2$ minimal model $(\hat{c}=k/(k+2))$ \cite{Dobrev,RY1}
are described 
as the branching functions of 
the Kazama-Suzuki coset \cite{KS} $\dsp \frac{SU(2)_k\times U(1)_2}{U(1)_{k+2}}$
defined by
\begin{eqnarray}
&& \chi_{\ell}^{(k)}(\tau,w)\Th{s}{2}(\tau,w-z)
=\sum_{\stackrel{m\in \bsz_{2(k+2)}}{\ell+m+s\in 2\bsz}} \chi_m^{\ell,s}
(\tau,z)\Th{m}{k+2}(\tau,w-2z/(k+2))~, \nn
%&& \hspace{6cm} \mbox{for $\ell+m+s \in 2\bz$}~, \nn
&& \chi^{\ell,s}_m(\tau,z) \equiv  0~, ~~~ \mbox{for $\ell+m+s \in 2\bz+1$}~,
\label{branching minimal}
\end{eqnarray}
where $\chi_{\ell}^{(k)}(\tau,z)$ is the spin $\ell/2$ character of 
$SU(2)_k$;
\begin{eqnarray}
&&\chi^{(k)}_{\ell}(\tau, z) 
=\frac{\Th{\ell+1}{k+2}(\tau,z)-\Th{-\ell-1}{k+2}(\tau,z)}
                        {\Th{1}{2}(\tau,z)-\Th{-1}{2}(\tau,z)}
\equiv \sum_{m \in \bsz_{2k}}\, c^{(k)}_{\ell,m}(\tau)\Th{m}{k}(\tau,z)~.
\label{SU(2) character}
\end{eqnarray}
The branching function $\chi^{\ell,s}_m(\tau,z)$ 
is explicitly calculated as follows;
%(see, {\em e.g.} \cite{KYY});
\begin{equation}
\chi_m^{\ell,s}(\tau,z)=\sum_{r\in \bsz_k}c^{(k)}_{\ell, m-s+4r}(\tau)
\Th{2m+(k+2)(-s+4r)}{2k(k+2)}(\tau,z/(k+2))~.
\end{equation}
Then, the character formulas of unitary representations 
are written as 
\begin{eqnarray}
&& \ch{(\sNS)}{\ell,m}(\tau,z) = \chi^{\ell,0}_m(\tau,z)
+\chi^{\ell,2}_m(\tau,z),
%\label{minimal character}
\nn
&& \ch{(\stNS)}{\ell,m}(\tau,z) = \chi^{\ell,0}_m(\tau,z)
-\chi^{\ell,2}_m(\tau,z)
\nn
%\equiv 
%e^{-i\pi\frac{m}{k+2}}\ch{(\sNS)}{\ell,m}\left(\tau,z+\frac{1}{2}\right)~, \nn
&& \ch{(\sR)}{\ell,m}(\tau,z) = \chi^{\ell,1}_m(\tau,z)
+\chi^{\ell,3}_m(\tau,z) 
\nn
%\equiv 
%q^{\frac{k}{8(k+2)}}y^{\frac{k}{2(k+2)}}
%\ch{(\sNS)}{\ell,m+1}\left(\tau,z+\frac{\tau}{2}\right)~, \nn
&& \ch{(\stR)}{\ell,m}(\tau,z) = \chi^{\ell,1}_m(\tau,z)
-\chi^{\ell,3}_m(\tau,z) .
%\equiv
%- e^{-i\pi\frac{m+1}{k+2}}q^{\frac{k}{8(k+2)}}y^{\frac{k}{2(k+2)}}
%\ch{(\sNS)}{\ell,m+1}\left(\tau,z+\frac{1}{2}+\frac{\tau}{2}\right)~. \nn
%&& 
\label{minimal character}
\end{eqnarray}
%By definition, we may restrict to $\ell+m \in 2\bz$ for the  $\NS$ and
%$\tNS$ sectors, and to $\ell+m \in 2\bz+1$ for the $\R$ and $\tR$
%sectors.  It is convenient to define 
%$\ch{(\sigma)}{*}(\tau,z)\equiv 0$ unless these conditions for $\ell$,
%$m$ are satisfied.

~

%%%%%%%%%%%%%%%%%%%%%%%%%%%%%%%%%%%%%%%%%%%%%%%%%%%%%%%%%%%%%%%%%%%%
%%%%%%%%%%%%%%%%%%%%%%%%%%%%%%%%%%%%%%%%%%%%%%%%%%%%%%%%%%%%%%%%%%%%
%%%%%%%%%%%%%%%%%%%%%%%%%%%%%%%%%%%%%%%%%%%%%%%%%%%%%%%%%%%%%%%%%%%%

\section*{Appendix B: ~ Explicit Forms of Spectral Flow Orbits and Their Orbifold Twistings}

\setcounter{equation}{0}
\def\theequation{B.\arabic{equation}}

~

In Appendix B, we summarize the explicit expressions of 
spectral flow orbits \eqn{cFNS}-\eqn{cFtR}, which play the role of building blocks of relevant modular invariants, 
and their twistings by the orbifold actions $\gamma_L$, $\delta_R$
introduced in section \ref{subsec:orbifold actions}.

%%%%%%%%%%%%%%%%%%%%%%%%%%%%%%%%%%%%%%%%%%%%

We make use of the abbreviated index $I \equiv \left\{(\ell_i, m_i) \right\}$ ($\ell_i+m_i \in 2\bz$) again, and 
set 
\begin{equation}
Q(I) \equiv Q \left(\left\{(\ell_i, m_i) \right\}\right) := \sum_{i=1}^r \, \frac{m_i}{k_i+2} \left(\in \frac{1}{N}\bz\right),
\end{equation}
for the convenience. 
%$\dsp Q(I) \in \frac{1}{N}\bz$
$\cF^{(\sigma)}_I(\tau,z) $ obviously vanishes  for $Q(I) \not\in \bz $ by the definitions \eqn{cFNS}-\eqn{cFtR}, 
and  
we obtain the following expressions in the case of $Q(I) \in \bz$,  
%%%%%%%%%%%%%%%%%%%%%%%%%%%%%%%%%%%%%%%%%%%%%%%%%%%%%%%%%%
\begin{align}
\cF^{(\sNS)}_I (\tau,z) & = \sum_{n\in\bz_N}\, \prod_{i=1}^r\, \ch{(\sNS)}{\ell_i, m_i - 2n}(\tau,z)
%%%%
\equiv \sum_{n\in\bz_N}\, F^{(\sNS)}_{s_n(I)}(\tau, z),
%%%%
\label{cFNS 2}
\\
\cF^{(\stNS)}_I (\tau,z) & = (-1)^{Q(I)} \sum_{n\in\bz_N}\, (-1)^{\left(\hc-r\right) n} \prod_{i=1}^r\, \ch{(\stNS)}{\ell_i, m_i - 2n}(\tau,z)
%%%%
\equiv \sum_{n\in\bz_N}\, (-1)^{\left(\hc-r\right) n} F^{(\stNS)}_{s_n(I)}(\tau, z),
%%%%
\label{cFtNS 2}
\\
\cF^{(\sR)}_I (\tau,z) & = \sum_{n\in\bz_N}\, \prod_{i=1}^r\, \ch{(\sR)}{\ell_i, m_i - 2n-1 }(\tau,z)
%%%%
\equiv \sum_{n\in\bz_N}\, F^{(\sR)}_{s_n(I)}(\tau, z),
%%%%
\label{cFR 2}
\\
\cF^{(\stR)}_I (\tau,z) & = (-1)^{Q(I)+r} \sum_{n\in\bz_N}\, 
(-1)^{\left(\hc-r\right) n} \prod_{i=1}^r\, \ch{(\stR)}{\ell_i, m_i - 2n-1}(\tau,z)
\equiv \sum_{n\in\bz_N}\, (-1)^{\left(\hc-r\right) n} F^{(\stR)}_{s_n(I)}(\tau, z),
\label{cFtR 2}
\end{align}
where we introduced the notation
$$
s_n(I) := \left\{ (\ell_i, m_i -2n)\right\} ~~  
\left( \mbox{for} ~  I \equiv \left\{ (\ell_i, m_i)\right\}\right) .
$$

%%%%%%%%%%%%%%%%%%%%%%%%%%%%%%%%%%%%%%%%%%%%%%%%%%%%%%%%%%
Then, the explicit actions of $\gamma$ and $\delta$-twisting\footnote
  {Here we omit the subscripts `$L$' and `$R$' used in the main text.} onto 
$\cF^{(\sigma)}_I(\tau)$ are evaluated as follows;
($\cF^{(\sigma)}_I(\tau) \equiv \cF^{(\sigma)}_I(\tau, 0)$, $a,b \in \bz_2$, $\al, \beta \in \bz_4$, $k_1+2 =4K \in 4\bz_{>0}$)
%%%%%%%%%%%%%%%%%%%%%%%
% action of $\gamma$
%%%%%%%%%%%%%%%%%%%%%%%
\begin{align}
\gamma_{(a,b)}\cdot \cF^{(\sigma)}_{\left\{(\ell_i, m_i) \right\}} (\tau)
= \left\{
\begin{array}{ll}
(-1)^{b\ell_1} \cF^{(\sigma)}_{\left\{(\ell_i, m_i)\right\}}(\tau), & ~~ (a =0)
\\
(-1)^{b(\ell_1+1)} \cF^{(\sigma)}_{\left\{(k_1-\ell_1, m_1), (\ell_2, m_2), \ldots, (\ell_r, m_r) \right\}}(\tau), & ~~ (a =1)
\end{array}
\right.
\label{gamma cF}
\end{align}
%%%%
%%%%%%%%%%%%%%%%%%%%%%%%
% action of $\delta$
%%%%%%%%%%%%%%%%%%%%%%%%
\begin{align}
\delta_{(\al,\beta)} \cdot \cF^{(\sigma)}_{\left\{(\ell_i, m_i) \right\}} (\tau)
& = \zeta_{K-\frac{1}{2}}(\sigma; \al , \beta)\, 
e^{2\pi i \frac{K}{4}(2K-1) \al \beta} 
\nn
& \hspace{1cm} \times  \sum_{n\in \bz_N}\, e^{2\pi i \frac{m_1-2n}{4}\beta} \, 
F^{(\sigma)}_{\left \{(\ell_1, m_1 -2n - 2K  \al), (\ell_2, m_2-2n) , \ldots, 
(\ell_r, m_r-2n) \right\}} (\tau),
\label{delta cF} 
\end{align}
where we introduced the phase factor
\begin{equation}
\zeta_{\kappa}(\NS; \al, \beta) =1, ~~\zeta_{\kappa}(\tNS; \al, \beta) =e^{-i\pi \kappa\al}, ~~ 
\zeta_{\kappa}(\R; \al, \beta) = e^{i\pi \kappa \beta}, ~~
\zeta_{\kappa}(\tR; \al, \beta) = e^{-i\pi \kappa(\al-\beta)}. 
\label{zeta kappa}
\end{equation}

~

%%%%%%%%%%%%%%%%%%%%%%%%%%%%%%%%%%%%%%%%%%%%%%%%%%%%%%%%%%%%%%%%%%%%
%%%%%%%%%%%%%%%%%%%%%%%%%%%%%%%%%%%%%%%%%%%%%%%%%%%%%%%%%%%%%%%%%%%%

For the ones given in section \ref{generalization}, we can also summarize as follows;
%%%%%%%%%%%%%%%%%%%%%%%%%%%%%%%%%%%
%%%%%%%%%%%%%%%%%%%%%%%
% action of $\gamma$
%%%%%%%%%%%%%%%%%%%%%%%
\begin{align}
\gamma_{(a,b)}\cdot \cF^{(\sigma)}_{\left\{(\ell_i, m_i) \right\}} (\tau)
= \left\{
\begin{array}{ll}
(-1)^{b\sum_{i\in S_1}\ell_i} \cF^{(\sigma)}_{\left\{(\ell_i, m_i) \right\}}(\tau), & ~~ (a =0)
\\
(-1)^{b\sum_{i\in S_1}(\ell_i+1)} \cF^{(\sigma)}_{\left\{(\ell'_i, m_i) \right\}}(\tau), & ~~ (a =1)
\end{array}
\right.
\label{gamma cF S}
\end{align}
%%%%
where we set
$$
\ell'_i : = \left\{
\begin{array}{ll}
k_i- \ell_i & ~~ i \in S_1,
\\
\ell_i & ~~ \mbox{otherwise}.
\end{array}
\right.
$$
%%%%%%%%%%%%%%%%%%%%%%%%
% action of $\delta$
%%%%%%%%%%%%%%%%%%%%%%%%
\begin{align}
\delta_{(\al,\beta)} \cdot \cF^{(\sigma)}_{\left\{(\ell_i, m_i) \right\}} (\tau)
& = \zeta_{\hc_S L}(\sigma; \al , \beta)\, e^{2\pi i \frac{L^2M}{N'} \al \beta} 
\sum_{n\in \bz_N}\, e^{2\pi i \sum_{i\in S}\, \frac{L \left(m_i-2n\right)}{k_i+2}\beta} \, 
F^{(\sigma)}_{\left\{(\ell_i, m''_i-2n)\right\}} (\tau)
\nn
& \equiv \zeta_{2LM/N'}(\sigma; \al , \beta)\, e^{2\pi i \frac{L^2M}{N'} \al \beta} 
\sum_{n\in \bz_N}\, e^{2\pi i \frac{L}{N'} \sum_{i\in S}\, d_i (m_i-2n)  \beta} \, 
F^{(\sigma)}_{\left\{(\ell_i, m''_i-2n )\right\}} (\tau),
\nn
& 
\hspace{8cm}
\left(d_i \equiv \frac{N'}{k_i+2} \right),
\label{delta cF S} 
\end{align}
where we set
$$
m''_i : = \left\{
\begin{array}{ll}
m_i - 2L \al & ~~ i \in S,
\\
m_i & ~~ \mbox{otherwise}.
\end{array}
\right.
$$

~

%%%%%%%%%%%%%%%%%%%%%%%%%%%%%%%%%%%%%%%%%%%%%%%%%%%%%%%%%%%%%%%%%%%%%
%%%%%%%%%%%%%%%%%%%%%%%%%%%%%%%%%%%%%%%%%%%%%%%%%%%%%%%%%%%%%%%%%%%%%
%%%%%%%%%%%%%%%%%%%%%%%%%%%%%%%%%%%%%%%%%%%%%%%%%%%%%%%%%%%%%%%%%%%%%
%%%%%%%%%%%%%%%%%%%%%%%%%%%%%%%%%%%%%%%%%%%%%%%%%%%%%%%%%%%%%%%%%%%%%
%%%%%%%%%%%%%%%%%%%%%%%%%%%%%%%%%%%%%%%%%%%%%%%%%%%%%%%%%%%%%%%%%%%%%

\end{document}